\documentclass[submission,copyright,creativecommons]{eptcs}
\providecommand{\event}{MARS 2015} 
\usepackage{amsmath}
\usepackage[pdftex]{graphicx}
\usepackage{epstopdf}
\usepackage{fancyvrb}
\usepackage{latexsym}
\usepackage{amssymb}
\usepackage{float}
\usepackage{multicol}
\usepackage{url}

\usepackage{breakurl}

\def\bitcoinA{%
  \leavevmode
  \vtop{\offinterlineskip 
    \setbox0=\hbox{B}%
    \setbox2=\hbox to\wd0{\hfil\hskip-.03em
    \vrule height .3ex width .15ex\hskip .08em
    \vrule height .3ex width .15ex\hfil}
    \vbox{\copy2\box0}\box2}}
\floatstyle{ruled}
\newfloat{program}{thp}{lop}
\floatname{program}{Declaration}
\newfloat{program2}{thp}{lop}
\floatname{program2}{Property}
\fvset{gobble=0,numbers=left,numbersep=3pt,stepnumber=5}
\begin{document}
\title{Modeling and Verification of the Bitcoin Protocol}
\author{Kaylash Chaudhary 
\institute{School of Computing, Information \\and Mathematical Sciences\\
University of the South Pacific\\Fiji
 \email{chaudhary\underline{ }k@usp.ac.fj}} \and Ansgar Fehnker
\institute{School of Computing, Information \\and Mathematical Sciences\\
University of the South Pacific\\Fiji\email{fehnker\underline{ }a@usp.ac.fj}}\\\quad\qquad\and Jaco van de Pol
\institute{Formal Methods and Tools\\
University of Twente\\
The Netherlands
}
\email{\quad j.c.vandepol@utwente.nl }\and Marielle Stoelinga
\institute{Formal Methods and Tools\\
University of Twente\\
The Netherlands
\email{ m.i.a.stoelinga@cs.utwente.nl}
}
 }
\def\titlerunning{Modeling and Verification of the Bitcoin Protocol}
\def\authorrunning{K. Chaudhary, A. Fehnker, J.C. van de Pol \& M.I.A. Stoelinga} 
\def\copyrightholders{Chaudhary, Fehnker, van de Pol \& Stoelinga}
\maketitle
\thispagestyle{empty}
\newcommand{\spending}{\textsc{Spending}}
\newcommand{\idleloc}{\textsc{Idle}}
\newcommand{\chargeloc}{\textsc{Charge}}
\newcommand{\waitloc}{\textsc{Wait}}
\newcommand{\transloc}{\textsc{Transfer}}
\newcommand{\buycoin}{\textsc{BuyCoin}}

\newcommand{\nocoin}{\textsc{NoCoin}}
\newcommand{\haveloc}{\textsc{HaveLoc}}
\newcommand{\reqloc}{\textsc{ReqLoc}}
\newcommand{\reqcoin}{\textsc{ReqCoin}}
\newcommand{\recvreq}{\textsc{RecvReq}}
\newcommand{\havecoin}{\textsc{HaveCoin}}

\newcommand{\failedloc}{\textsc{Failed}}

\begin{abstract}
Bitcoin is a popular digital currency for online payments, realized as
a decentralized peer-to-peer electronic cash system. Bitcoin keeps a
ledger of all transactions; the majority of the participants decides on
the correct ledger. Since there is no trusted third party to guard
against double spending, and inspired by its popularity, we would like to
investigate the correctness of the Bitcoin protocol. Double spending
is an important threat to electronic payment systems. Double spending
would happen if one user could force a majority to believe that a
ledger without his previous payment is the correct one. We are
interested in the probability of success of such a double spending
attack, which is linked to the computational power of the
attacker. This paper examines the Bitcoin protocol and provides its
formalization as an UPPAAL model. The model will be used to show how
double spending can be done if the parties in the Bitcoin protocol
behave maliciously, and with what probability double spending occurs.
\end{abstract}

\section{Introduction}

Bitcoin \cite{Nakamoto}  is a popular digital currency with over 81,398,896 transactions  and 14,544,775 bitcoins in circulation until August, 2015 \cite{Bitcoinchart}. Due to its popularity, many merchants have started accepting bitcoins, such as a coffee shop located on the campus of one of the universities in Mexico \cite{Bitcoincoffeeshop}. Bitcoin eliminates the need for a trusted third party, such as a broker or a bank, to process payments. Every peer in the Bitcoin network keeps the collection of all transactions.
This so-called ledger is organized in separate blocks, which are linked to their predecessor, thus forming a chain. Each peer maintains its own chain.  This chain will work effectively for honest peers.

A system is not only composed of honest peers but also dishonest ones. If the peers deviate from their expected behaviour, then the chain may fork and double spending may be possible. Recently, there was a fork in the network that created two versions of the block-chain and it continued for six blocks \cite{Doublespending}. That created the risk of double spending. Another reason that could increase the risk of double spending is the size of the pools. In June, 2014, Ghash.io, one of the mining pools of the Bitcoin network, came close to obtaining 51\% of the Bitcoin network's hash-rate \cite{doublespendingattack}. This could have led to double spending if the pool would have behaved maliciously.

Bitcoin attracted attention of researchers who worked on different aspects of the protocol. Bergstra et.\ al.\ identified research questions related to Bitcoin and other information money  \cite{DBLP:journalscorrabs}. Ron et.\ al.\ did quantitative analysis of the Bitcoin transaction graph \cite{DBLP:journals/iacr/RonS12}. Herrmann did the implementation, evaluation and detection of a double-spending attack \cite{herrmann2012implementation}. The Bitcoin protocol has been modeled before to analyze different properties. Beukema developed a model in mCRL2 and verified properties such as double spending \cite{Beukema}. This model consisted only of peers that are responsible for mining as well. Block-chain forking was not investigated. Andrychowicz et.\ al.\ modeled Bitcoin contracts using timed automata \cite{Andrychowicz}. Bastiaan did stochastic analysis of two-phase proof-of-work in the Bitcoin protocol \cite{Bastian}. This work was based on predicting the rate at which miners generate coins. However, it did not include block-chain forking. Our research focuses on block-chain forking and the probability of double spending, taking into account that the block-chain can be influenced by a malicious pool.

This paper focuses on the formal modeling of the Bitcoin protocol, developed in the UPPAAL model checker. It also presents the preliminary results of our probability analysis for double-spending attacks, based on the number of confirmations. This model will be used in the future to further analyze double-spending attacks, based on the computational power of the pools.

\paragraph{Organization of the paper.}
The next section introduces the Bitcoin protocol. A discussion on the model is presented in Section~\ref{discussion}. Appendix~\ref{description} shows the full description of the Bitcoin protocol in UPPAAL. Our basic analysis of this protocol is discussed in Section~\ref{correctness}. The paper concludes with a discussion on future possibilities for analysis, based on the developed model, in Section~\ref{conclusion}.


\section{The Bitcoin Protocol}\label{bitcoin}

The Bitcoin protocol proposed by Nakamoto is described as a peer-to-peer electronic cash system \cite{Nakamoto}. The Bitcoin protocol is a decentralized payment system. There is no trusted third party, such as a bank or a broker, as used in traditional  payment systems to guard against double spending. In this online payment system, a payer sends a payment directly to a payee. Bitcoin is a digital currency based on cryptographic principles. Every peer knows about every other peer's transactions; this is unlike a centralized system (Bank) where only the central authority knows about all transactions. Each peer in the Bitcoin network can be a payer (customer) or payee (vendor) or both. To start using the Bitcoin protocol, the user needs a Bitcoin account and a wallet. A prerequisite for a user is to have a private and a public key pair, first of all because the account is identified by the public key. Also, a payer needs to know the public key of the payee to make a payment. The next subsection will discuss the basic terminology.

\subsection{Terminology}\label{terminologies}
The following cryptography and payment terminology is used by the Bitcoin protocol \cite{Antonopoulos:2014:MBU:2695500}.
\begin{itemize}
    \item \textbf{One-way hash function}: Bitcoin uses the double-SHA256 hashing algorithm to hash transactions and to solve proof-of-work puzzles.
    \item \textbf{E-wallet}: An e-wallet is a database that stores bitcoins and their corresponding transaction hashes. It is usually referred to as the bitcoin wallet which can be software, mobile or web-based.
    \item \textbf{Bitcoin}: Bitcoin is a digital currency exchanged between parties over the Bitcoin network. BTC or \bitcoinA  \;is the currency symbol for bitcoin.
    \item \textbf{Proof-of-work}: A cryptographic puzzle (also known as a challenge) is based on Adam Back's Hashcash \cite{Back02hashcash-}. Originally, Hashcash was developed as an anti-spam  email tool, where stamps are created and attached to the email. A random number is added to the email header and then a hash is performed on the header. A hash string is considered \emph{valid} if it has a value equal to some set target. In case of an invalid hash string, another random number is selected, added to the header and the hash is computed again. This process continues until a valid hash string is obtained. The recipient checks for the validity of the date and email address in the hash string and adds the hash string into the database. If the hash string is already present in the database, that hash string is invalid.  Bitcoin uses Hashcash for block validity too. Miners perform proof-of-work calculations on recently broadcasted transaction blocks. This process is called \emph{mining} and is performed by mining software. Mining yields bitcoins for the miner as a reward.
    \item \textbf{Block-chain}: The block-chain is a public ledger which stores processed transactions.

\end{itemize}

If a user wants to make a payment, that user has to create a transaction with all relevant information, i.e., the address of the payee, the amount of bitcoins and a challenge. This transaction is then broadcasted to all nodes in the network. Each node places this transaction into a block and tries to solve the proof-of-work for this block. Once a node finds a solution to the challenge, it broadcasts it to all other nodes. If all transactions are valid and not spent before, then the other nodes accept the block and start working on the next block in the chain. At certain times, the chain may have forks because two blocks were mined and broadcasted at the same time. Nodes always consider the longest block-chain to be the correct one. The following subsections will discuss transactions, blocks and the block-chain in more detail.

\subsection{Transactions}

There are two types of transactions: coin-base and regular transactions \cite{bitcoindeveloper}. Coin-base transactions are used for new bitcoins, whereas regular transactions are used for transferring existing bitcoins from one user to another. Each transaction has one or more transaction inputs and outputs. A \emph{transaction input} is a reference to an output of the previous transaction, which proves that the sender possesses the bitcoins it claims to have. A transaction input contains the response corresponding to the challenge of the previous transaction output. A \emph{transaction output} determines the amount to be transferred to a payee account. Transaction outputs consist of an amount in BTC and a challenge, which specifies the conditions under which this output transaction can be claimed.

\subsection{Blocks}

A block is a set of transactions. Transactions in the same block can be considered to have happened at the same time. Transactions are only confirmed if they appear in some block. Unconfirmed transactions are kept in a transaction pool.  Any node can create a block of transactions and broadcast it to the rest of the network. Since there can be many block creators, the network needs to decide which block should be added to the chain next, as the blocks can arrive in a different order to different peers. The protocol uses the proof-of-work solution to induce a unique order on blocks.

\begin{figure}[t!]
  \centering
  \includegraphics[width = 8.5cm]{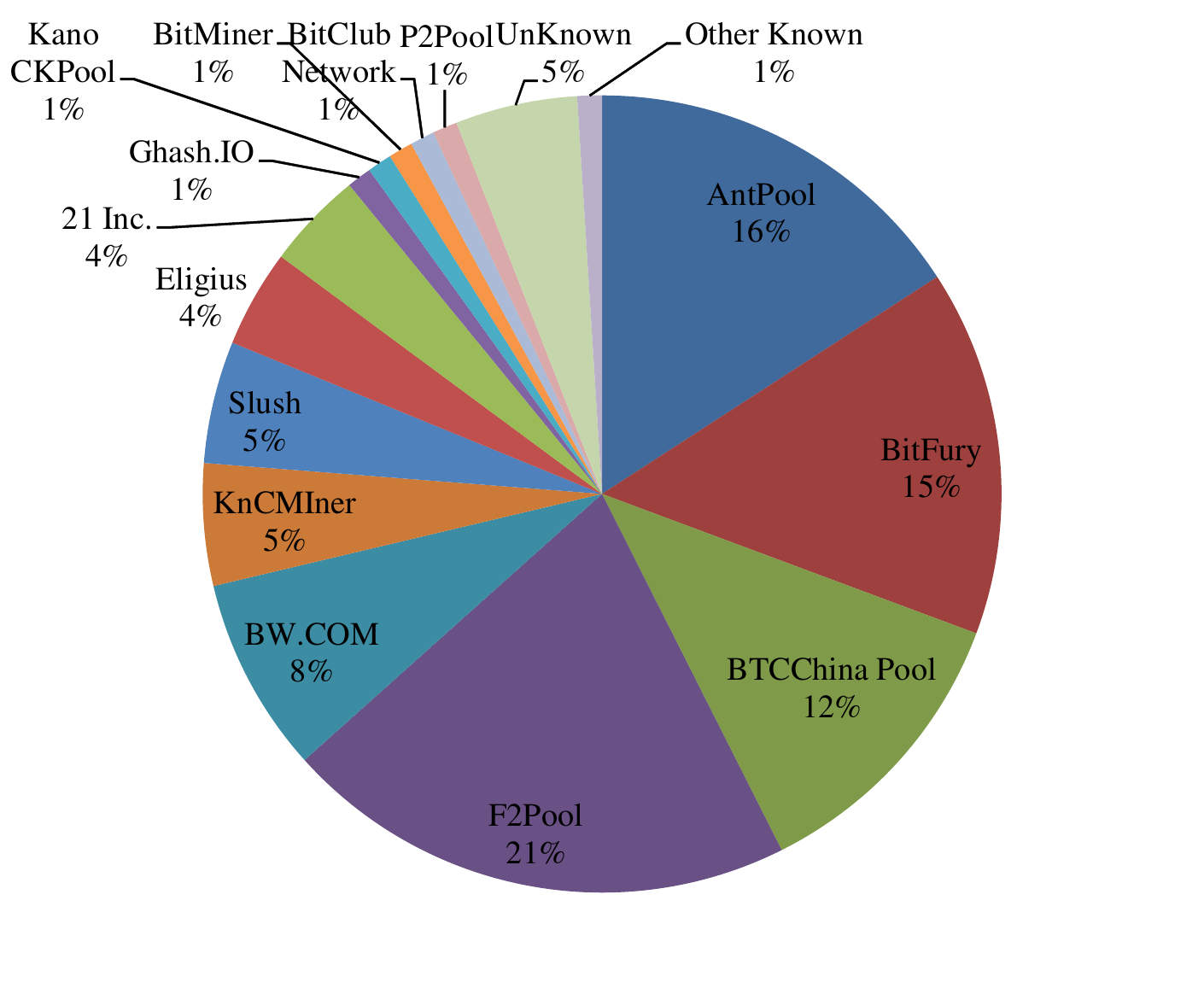}\\
  \caption{Pools with percentage hash-rate   }\label{poolchart}
\end{figure}

This proof-of-work solution is accomplished by the mining process. Mining is done by peers, called miners, or even by pools of miners. Initially, ordinary peers could participate in the mining process. Since the difficulty for finding a solution to the proof-of-work increased over the years, peers started working together in groups to solve the challenges. Such a group is called a pool. Today, there are many pools available with different percentages of participation in the Bitcoin network as shown in Figure \ref{poolchart} (August, 2015)\cite{hashrate}. Pool $\textit{BitFury}$ occupies 15\% of the network hash-rate.
The network hash-rate (hashes per second) is a measure for the processing power of the Bitcoin network.  Related to the network hash-rate, the Bitcoin network has a global block difficulty which is automatically adapted after every 2016 blocks. It is a measure to ensure that it is not too easy to solve the challenges with the current network hash-rate. This difficulty is based on a target value: the miners must find a block solution less than the target value. So the target value is a maximum hash. A decrease in target value means an increase in difficulty. On 1st August, 2015, the network hash rate was 413,204,212.12 GH/s and the difficulty was 52,278,304,845.59 \cite{hashrateAndDifficulty}. The average time taken to find a solution by a particular pool is based on the hash-rate of that pool. The time is calculated using Equation \ref{time} \cite{Bastian}.

\begin{equation} \label{time}
\begin{split}
{\textit Time} & = \frac{ \textit{difficulty}*2^{32}}{\textit{hash-rate}}
\end{split}
\end{equation}

For example, the average time taken to find a block solution by \emph{BitFurry }is:
\begin{equation} \label{timeBitFurry}
\begin{split}
{\textit Time}
    & = \frac{ 52,278,304,845.59*2^{32}}{0.15*(413,204,212.12)}\\
    & = 3622\; s \simeq 60\;mins
\end{split}
\end{equation}

To solve a challenge, miners in a pool randomly select a so-called \emph{nonce} for a block. The hash of the block (including this nonce) is calculated and the result is compared with the target value. If the hash of the block is lower than the target value, then the proof-of-work is solved; otherwise the miners randomly select another nonce and calculate the hash again. This process repeats until a nonce is found that results in a hash value lower than the target value. Once a nonce is found, the block becomes valid and is broadcasted. On average, it takes about ten minutes for a pool to find a solution. Peers receiving this block can easily verify the solution by taking the hash of the block. Note, that the new block created by peers should include the hash of the most recent block on the longest path in the block-chain. Recall that every peer has its own version of the block-chain.
The first block, called the genesis block, is mined with the maximum target value.

The next section describes how blocks are used to form a chain.

\subsection{Block-Chain}
The block-chain is used to provide an order on the transactions, whereas the transaction chain is used to track the ownership history. Every valid block that is broadcasted joins the longest path of the chain, appending it to a predecessor in the longest chain. A block can only be added to the chain if it contains a hash of some previous block in the chain. A chain is formed by linking blocks which have a hash of a previous block in the block-chain. This defines an ordering of the blocks, thus preventing double spending.

A chain can have different paths but the longest path is the valid one.  Transactions in blocks that are not in the longest chain are added back to the pool of transactions, and they can be used to build new blocks. This process allows miners to select the newest block from the chain to create the next block. Selection of an older block can result in a shorter branch and making this branch grow longer than the current longest branch will require a lot of effort. Whenever a block becomes part of the block-chain, the miners are rewarded. The reward is included in every mined block as a coin-base transaction. This coin-base transaction is the first transaction in the block. This incentive will tempt miners to select the newest block.

Since bitcoins are used over a large network, in which peers and pools can even be temporarily disconnected, there is a possibility that two blocks are created and broadcasted over the network simultaneously. Some peers might receive the first block and other peers the second one. In this situation, peers will continue building the chain on the received block. This will be resolved when a new block will be broadcasted, received by all, since one of them makes the longest chain. This shows that the blocks which are at the end of the chain could be less secure, compared to the blocks at the start of the chain. The transactions in the block at the end of the chain should not be considered confirmed as there is a non-negligible possibility that another chain will become the longest.

Double spending is spending a coin twice. It is possible in Bitcoin if the block-chain can be influenced. There are different scenarios for double spending such as the race attack, Finney's attack, the brute force attack, and the $>50\%$ attack \cite{bitcoinattack}. The last attack relies on the hash-rate percentage of a pool. We are looking at double spending based on the hash-rate of pools, and how a pool can race in order to make its chain longer. This scenario is described in the next section.


\section{Bitcoin Model Discussion}\label{discussion}
We have used the UPPAAL statistical model checker to model and analyze the Bitcoin protocol \cite{DBLP:journals/sttt/DavidLLMP15}. The model consists of four types of automata: pool, peer, malicious peer and malicious pool. A peer creates transactions and places them in the transaction pool. A pool gets transactions from the transaction pool, includes them in a block and proceeds to the mining step. Upon successful mining, the mined block is broadcasted to other peers and pools for inclusion in the block-chain.

A malicious peer and pool collaborate together to cheat the system. The malicious peer creates two transactions of the same input, places one in the transaction pool for other pools to mine and gives the other one to the malicious pool. After the first transaction has been included in the chain, the malicious pool adds the block with the malicious transaction to the chain,  creating a fork. The race is between these two chains. If the chain with the malicious transaction exceeds the length of the chain with the first transaction then the former becomes valid and all pools continue building blocks on the wrong chain. The detailed model description is presented in Appendix~\ref{description}. This model was built with the intention to investigate the probability of double spending in relation to the pool size. We have abstracted the size of the pool by the rate at which blocks are generated by each pool. This rate is calculated using Equation \ref{time} in Section \ref{bitcoin}.

\begin{figure}[t!]
  \centering
  \includegraphics[width = 6cm]{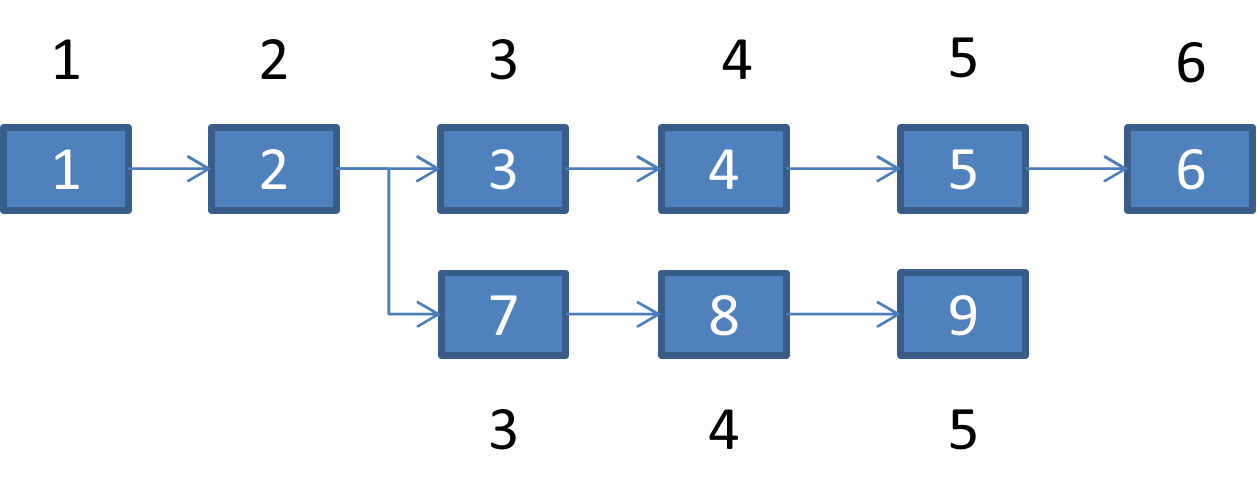}\\
  \caption{Block-chain fork with identifiers and length indicators}\label{blockchainlength}
\end{figure}

It was difficult to model the block-chain in UPPAAL due to the limited support for unbounded data structures.  A fork in the block-chain leads to a side chain. The longest chain is the main chain. A side chain could increase in length as well and possibly become the main chain later on. Therefore, it was important for us to store all chains. We modeled a block-chain by a list of triples, which store the current block number, the previous block number and the length of the chain to that block. For example, the following is the block-chain structure for the block-chain shown in Figure \ref{blockchainlength}.
\begin{verbatim}
{[1,0,1],[2,1,2],[3,2,3],[4,3,4],[5,4,5],[6,5,6],[7,2,3],[8,7,4],[9,8,5]}
\end{verbatim}
In Figure \ref{blockchainlength}, the length for the chain to block number 9 is 5 and the length of the chain to block number 6 is 6. So the former is the side chain and the latter is the main chain. The length stored in the chain was used to determine the longest chain, to insert a block in the longest chain and to check whether a block is in the longest chain.

\begin{figure}[t!]
  \centering
  \includegraphics[width = 8cm]{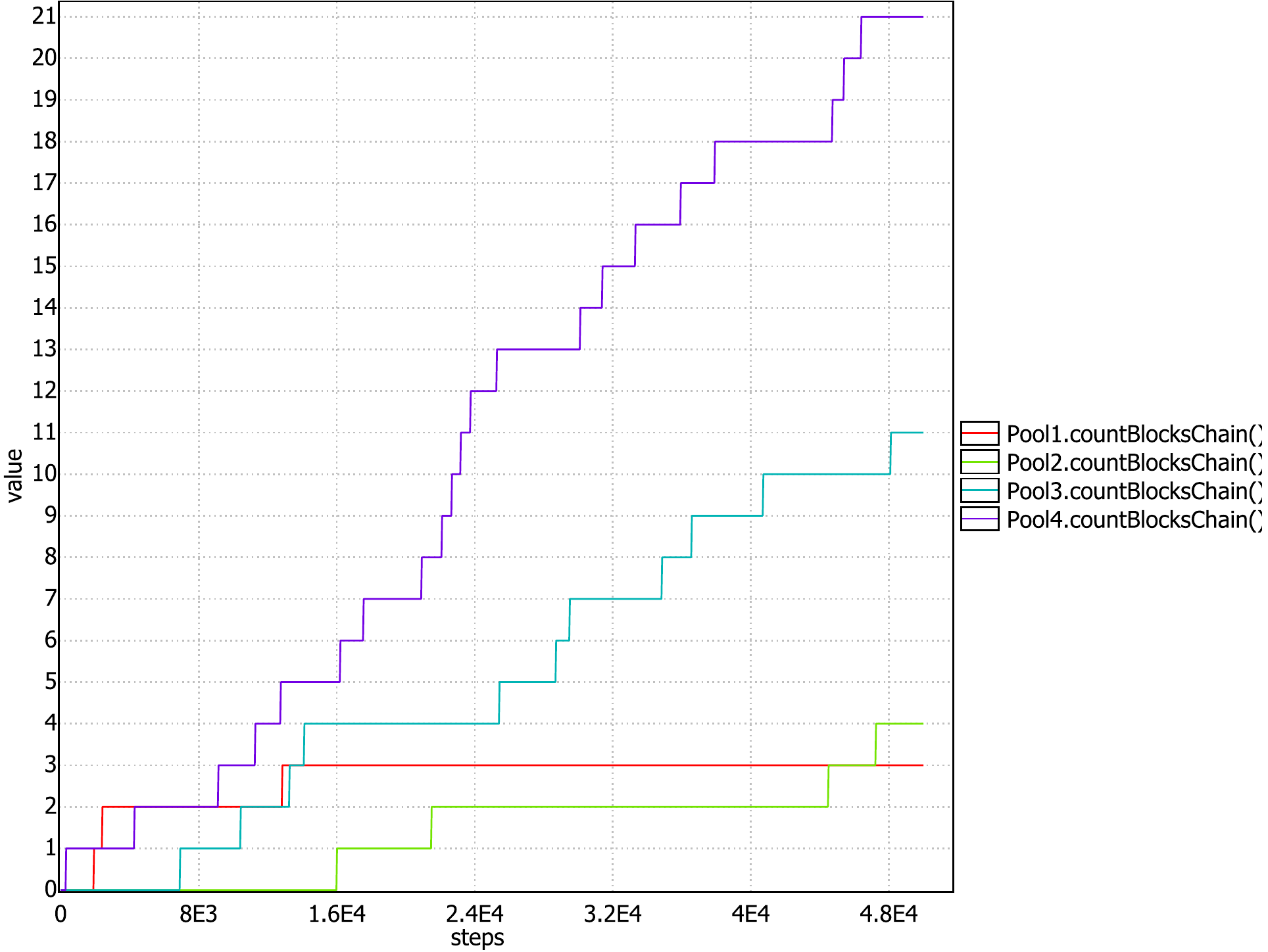}\\
  \caption{Number of blocks generated by each pool }\label{countBlock}
\end{figure}

To ensure the accuracy of the model, we simulated the model several times. We checked whether each functionality included in the model is working correctly and according to the Bitcoin protocol. One of the main functionalities includes validating a transaction before including it in a block. To test this functionality, we created a malicious peer who was responsible for sending negative or zero amounts, and already spent transactions. The model behaved as expected. Simulation was also done to validate the structure of the block-chain at each pool. Since we have abstracted the size of the pool by the hash-rate of the network, we simulated to validate whether the hash-rate corresponds to the number of blocks generated in the main chain.  In theory, a pool with a high hash-rate generates a larger number of blocks. Figure~\ref{countBlock} shows the result of this simulation.
Pool 4 has 50\% network hash-rate while Pools 1, 2 and 3  have 10\%, 18\% and 22\%, respectively. Our result validates that Pool 4 has created and successfully mined more blocks than the other pools.


\section{Basic Analysis of the Bitcoin Protocol}\label{correctness}
%
%
%
%

Double spending is possible in fast payment transactions according to Karame et.\ al.\ \cite{Karame:2012:DFP:2382196.2382292}. Recall that blocks in the main chain are less secure towards the end. Therefore, vendors should wait for a certain number of confirmations, before considering a transaction in a block confirmed. The number of confirmations is the depth of a block from the end of the chain. For a vendor to consider a transaction to be confirmed, it has been mentioned \cite{journals/iacr/BonneauMCNKF15} that the vendors need to wait for at least a depth of 6 blocks in the main chain. Based on this, we wanted to examine the relation between the hash-rate of a pool and the double-spending attack, following the scenario described in Section~\ref{discussion}.

We did the analysis on the possibility of double spending for depths of 1-4 blocks. The result presented in Table \ref{prob} is based on four peers and four pools. The hash-rate for the pools are 18\%, 22\% ,10\% and 50\%. The last one is the malicious pool. There are 3 honest peers and 1 malicious peer. The malicious peer collaborates with the malicious pool for the double-spending attack. The upper bound for the transactions and the blocks are set to 200. The query given below was used to get the probability of double spending by simulation:

\begin{verbatim}
            Pr[#<=500000](<> Pool4.checkBlockinChain(Pool4.MBlock))
\end{verbatim}

The malicious transaction by the malicious peer is included in the block $MBlock$ mined by Pool 4. The function $checkBlockinChain()$ checks whether this block is in the longest chain or not. If it is, the double spending has occurred. For each confirmation depth 1-4, this query was executed for five hundred thousand steps. 

\begin{table}[h]
\caption {Probability of double spending based on the confirmation depth} \label{prob}
\begin{center}
\begin{tabular}{|c|c|c|}
\hline No. of Confirmation  &  Probability of Double Spending & Runs\\\hline
 1   &[0.870781, 0.970278] & 118\\
2 &[0.855061, 0.954924] & 140  \\
3 &[0.797987, 0.897941] & 211 \\
4 &[0.734029, 0.833722] & 277\\
\hline
 \end{tabular}
 \end{center}
 \end{table}
Our results show that double spending is still possible with a depth of 4 blocks. The probability was approximately above 80\% for a depth of 4 blocks. The reason could be the hash-rate of the pool. However detailed analysis on double spending with different hash-rates is still needed.

%


\section{Conclusions and Future Research}\label{conclusion}

This paper presented a model of the Bitcoin protocol in the UPPAAL model checker and described how the important characteristics of the protocol have been modeled. The model was developed with the aim of investigating double spending in an environment where some peers are dishonest. This paper showed how a malicious pool could abuse its computational power and race against the network to include a malicious transaction in the longest chain. By statistical model checking, we analyzed how the probability that a malicious transaction is included in the longest chain, depends on the number of confirmations. The results suggest that the double-spending attack is possible for the parameters used in this scenario. However, our analysis is not complete yet, as we want to investigate also double spending with different hash-rates. 

Future research is to do rigorous analysis of the Bitcoin protocol using this model. One could investigate the malicious act of a pool under different network hash-rates. Also, it would be interesting to model the block-chain for the Bitcoin protocol with two-phase proof-of-work~\cite{Bastian}, and compare the probability of double spending with the standard Bitcoin protocol with single proof-of-work.


\bibliographystyle{eptcs}
\bibliography{references1}

\appendix

\section{Description of the Bitcoin Protocol in UPPAAL}\label{description}

This section provides the models and their description for the Bitcoin protocol. Block-chain, blocks and transactions form an integral component of the Bitcoin protocol. The validity of a transaction is dependent on the block-chain. If the block of a particular transaction is included in the main chain, then it can be considered confirmed. Due to its importance, we have modeled block-chains together with blocks and transactions. There are two parties involved: pools and peers. A Pool is a group of miners responsible for solving the proof-of-work problem for individual blocks. Each Pool maintains its own block-chain. A Peer can be a customer or a vendor who is responsible for transferring bitcoins from one peer to another by creating transactions. Peers do not take part in the mining process. This model will show the interaction between these two parties.

The next subsection will discuss the data structure used to store all necessary information (transactions and blocks) in this protocol.

\subsection{Global Declarations}\label{globaldeclaration}
Declaration \ref{transaction} shows the structure of a transaction. A transaction is made up  of an input and output transaction. Input transactions, $\textit{TXIN}$, contain the previous transaction. This means that this transaction is based on the output of the previous transaction (positive BTC). Output transactions, $\textit{TXOUT}$, contain the ${\textit Address}$ of the peer and the amount $\textit{value}$. Note that in the Bitcoin protocol a transaction does not have a sequence number. Instead, it has hash. We have abstracted this by a number in our model. To maintain the uniqueness of transactions, we have the transaction number $\textit{Tx\_num}$ stored and incremented globally. $\textit{Tx\_num}$ is initialized to $\textit{PEERS}$. This is because every peer will have one transaction to start off. Variable $\textit{status}$ is used to indicate the status of a transaction. It can be in three states; $\textit{UNCONFIRMED}$, $\textit{CONFIRMED}$ and $\textit{INVALID}$. Initially, a  transaction is $\textit{UNCONFIRMED}$. Once it is included in a block it is $\textit{CONFIRMED}$. If it can not be included in a block, its status will be $\textit{INVALID}$. $\textit{INVALID}$ is used to indicate to other pools that this transaction should not be included in a block. An invalid transaction has a negative or zero value in the output transaction, or it has already been spent.

Transactions are placed in a block. Declaration \ref{block} provides the structure of a block. Every block has a unique block number, a previous block number that forms a link to the main chain, and the transactions. A block also contains a hash.  We have represented it as a number $\textit{Block\_num}$ which is maintained globally. $\textit{Block\_num}$ is initialized to 2. The first new block created by a pool will be numbered 2. Block number 1 is reserved for the first (genesis) block.

\begin{program}[h]
\begin{multicols}{2}
 \begin{verbatim}
const int MAX_BTC_AMOUNT = 12;
const int MAX_TRANSACTION = 200;
const int PEERS = 4;
typedef int[0,PEERS] PEER;
typedef int[0,MAX_BTC_AMOUNT] AMOUNT;
typedef int[0,MAX_TRANSACTION]TXNUMBER;
typedef int[0,2] STATUS;

//output transaction
typedef struct{
    PEER Address;
    AMOUNT value;
}TXOUT;


//input transaction
typedef struct{
    TXNUMBER id;
}TXIN;

//transaction
typedef struct {
    TXNUMBER Id;
    TXIN TxInput;
    TXOUT TxOutput[2];
    STATUS status;	
}Tx;

TXNUMBER Tx_num = PEERS;
\end{verbatim}\vspace*{-2ex}
\caption{Transaction Datatype in UPPAAL}\label{transaction}
\end{multicols}

\end{program}

\begin{program}[h!]
\begin{multicols}{2}

 \begin{verbatim}
typedef int[0,HASHSIZE] BLOCKHASH;

typedef struct {
    BLOCKHASH num;
    BLOCKHASH prevBlocknum;	
    Tx tx;	
}Block;

BLOCKHASH Block_num = 2;
\end{verbatim}\vspace*{-2ex}
\caption{Block Datatype}\label{block}
\end{multicols}
\end{program}

We have also defined a structure $Wallet$ shown in Declaration \ref{wallet}. Every peer will have a wallet to maintain unspent output from every transaction.
\begin{program}[h!]

\begin{verbatim}
typedef struct{
    TXNUMBER Id;
    AMOUNT value;
}Wallet;
\end{verbatim}\vspace*{-2ex}
\caption{Wallet Datatype}\label{wallet}

\end{program}

Other global declarations include channels and variables. Channel $\textit{blockSolution}$ is used to broadcast blocks to other pools and peers. This channel is used when a pool has found a block solution. Channel $\textit{request}$ and $\textit{send}$ are used to request and send predecessors of orphan blocks. There are two meta variables used in this model. One is $\textit{Blocknum}$ used for storing block numbers when synchronizing on channel $\textit{request}$ and the second is $\textit{chain}$ used to store block details when broadcasting a block solution.

In the Bitcoin protocol, when a peer creates a transaction it broadcasts it over the network. A pool collects those transaction and places it in its pool. Instead of having a transaction pool for each pool, we have created a global transaction pool; $\textit{Tx\; TxPool[MAX\_TRANSACTION]}$.

The following subsections will discuss the two parties involved in the Bitcoin protocol.

\subsection{Pool}\label{pool}

\begin{figure}[t]
  \centering
  \includegraphics[width = 10cm]{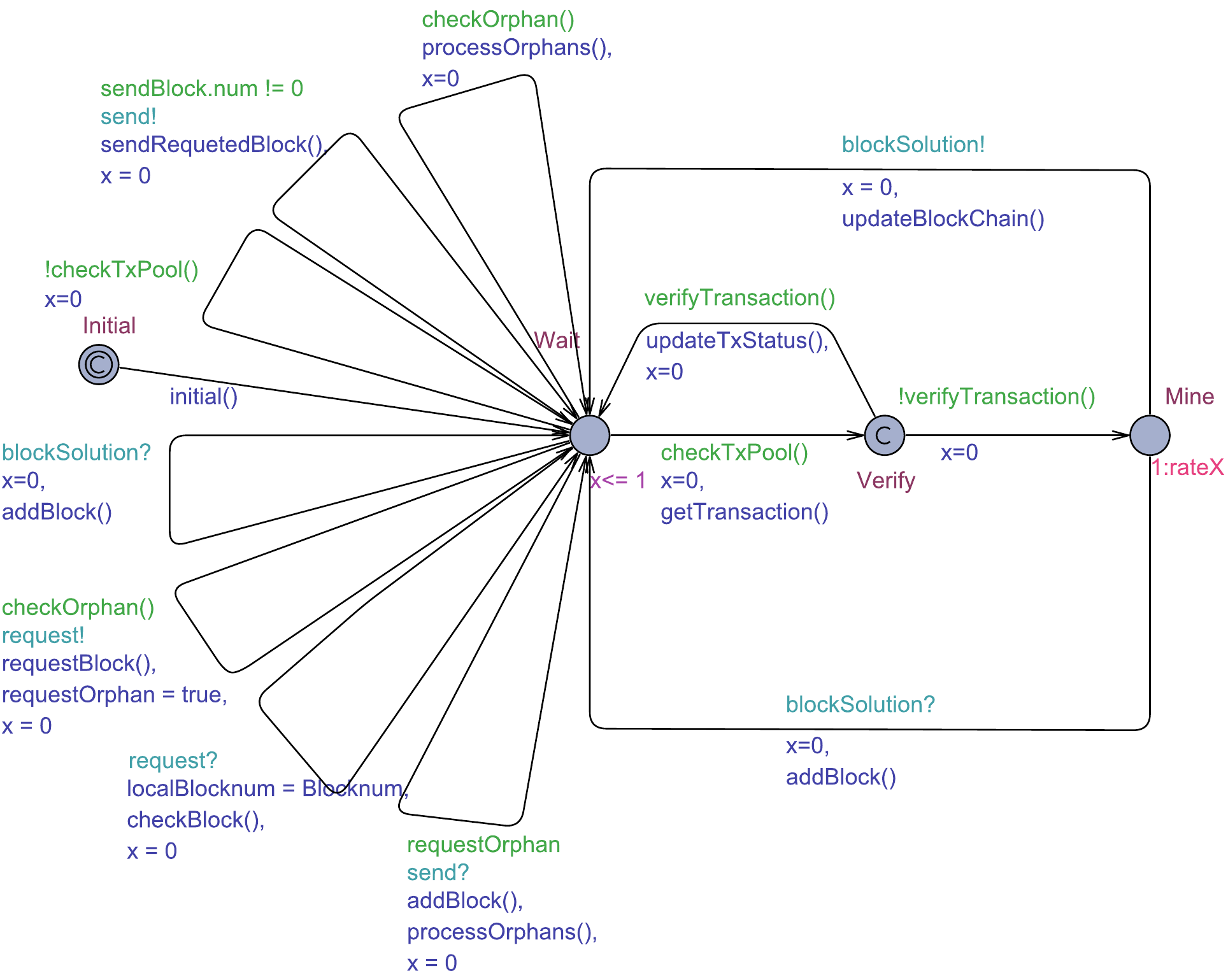}\\
  \caption{Pool automaton}\label{pooltemplate}
\end{figure}

Figure \ref{pooltemplate} shows the timed automaton $\textit{Pool}$. This automaton has four locations; $\textit{Initial}$, $\textit{Wait}$, $\textit{Verify}$ and $\textit{Mine}$. There are some local declarations, for the block-chain, the index of the longest chain and the length of the longest chain. We have implemented a block-chain as an array of integers, $\textit{chain[HASHSIZE][3]}$. This array records the block number, the previous block number, the length of the chain to that block, and the ID of this pool (in case this block was created by this pool). The reason behind storing only block numbers and the previous block number is to achieve simplicity of the chain description.  The third field in the array is used for the length of the chain.  The fourth field is used for verification purposes only, to determine the number of blocks created by a particular pool. The local variables $\textit{indexLongestChain}$ and $\textit{lengthLongestChain}$ are used to store the index and the length of the longest chain, respectively. These variables are used during block generation.
The block-chain contains only block numbers in our model. Other details included in a block, such as transactions, are stored in an array $\textit{confirmedBlock[MAX\_BLOCK]}$. Variable $\textit{confirmedBlock}$ stores all blocks mined by any pool. There can be orphan blocks in the protocol. Orphan blocks are those whose predecessor is not in any chain. Therefore, an array is defined for orphan blocks, $\textit{orphanBlock[MAX\_BLOCK]}$. Variables $\textit{localTx}$ and $\textit{localBlocknum}$ are used to store intermediate results when verifying transactions and processing orphans.

This automaton models the three tasks of the pool. The first task is verifying and placing a transaction in a block. The second is mining the block to solve the proof-of-work challenge. Upon successful mining, the block is broadcasted to other pools and peers. The final task is maintaining the block-chain by adding blocks received from other pools, and requesting the predecessors of orphans.

The automaton starts from location $\textit{Initial}$, taking a transition to $\textit{Wait}$. During the transition, the automata initializes the block chain $\textit{chain}$ and sets the first block in the chain to 1. So when the first block is created by this pool, it should refer to block 1 as the previous block. At the location $\textit{Wait}$, there are 5 outgoing edges and the rest are incoming edges, synchronized on channels with other automata.

If there are unconfirmed transactions in the transaction pool then the transition with the guard $\textit{checkTxPool()}$ will be enabled. The automaton will transit to location $\textit{Verify}$ if this transition is taken. The function $\textit{getTransaction()}$ copies the transaction into $\textit{localTx}$ and sets this transaction as confirmed in the transaction pool. The $\textit{verifyTransaction()}$ function checks the transaction for negative or zero values in the output section and it also compares the input of this transaction and the total value of the output transactions with the transactions in the main chain. The latter implements the double-spending check. If the input transaction number and the total value match with another transaction, then the input of this transaction has already been spent. Thus, the automaton transits to $\textit{Wait}$, setting the transaction status in the pool as $\textit{INVALID}$. If the check succeeds, then the automaton transits to the $\textit{Mine}$ location. The location $\textit{Mine}$ has the rate of exponential set to $\textit{1:rateX}$. This $\textit{rateX}$ is calculated from Equation \ref{time} in Section \ref{bitcoin}.

At this location, if a pool solves the proof-of-work challenge then it takes a transition to location $\textit{Wait}$ by broadcasting the block to other pools and peers on channel $\textit{blockSolution}$. The function $\textit{updateBlockChain()}$ adds the block with the block number stored in $\textit{Block\_num}$ and the transaction stored in the $\textit{localTx}$ to $\textit{confirmedBlock}$. This function also updates the global variable $\textit{chain}$, increments $\textit{Block\_num}$, resets $\textit{localTx}$ and adds the block number and the previous block number to its block-chain $\textit{chain}$.  The block found by the pool will always be added to the main chain and it will refer to the last block in the main chain. The index of the last block in the main chain is stored in variable $\textit{indexLongestChain}$.

If some other pool solves the proof-of-work challenge, then this pool will take the transition to the $\textit{Wait}$ location by receiving the block on channel $\textit{blockSolution}$. The function $\textit{addBlock()}$ checks if the received block's parent is in the chain and if it is, then the block is added to $\textit{confirmedBlock}$. The block number and the previous block number are added to the block-chain $\textit{chain}$. It also sets the transaction that it included in the block being mined to $\textit{UNCONFIRMED}$. If the parent is not in the $\textit{chain}$, then this block is an orphan and it is added to $\textit{orphanBlock}$. In either case, the transition is from the $\textit{Mine}$  to the $\textit{Wait}$ location. Likewise, if a pool is at the $\textit{Wait}$ location, it can also receive a block on channel $\textit{blockSolution}$.

If there is no transaction in the transaction pool, the transition with the guard $\textit{!checkTxPool}$ will be taken. If a pool has orphan blocks in the $\textit{orphanBlock}$, then the transition with the guard $\textit{checkOrphan()}$ will be enabled. This will let the pool request for the parent of the orphan block on channel $\textit{request}$. The function $\textit{requestBlock()}$ updates the global variable $\textit{blocknum}$ with the block number of the parent. The automaton synchronizing on channel $\textit{request}$ will copy the value of $\textit{Blocknum}$ to $\textit{localBlocknum}$. It will then check whether a block with block number $\textit{localBlocknum}$ is present in the $\textit{confirmedBlock}$. If it is, then the block information is saved in the local variable $\textit{sendBlock}$. When $\textit{sendBlock}$ has information in it, the transition with the condition $\textit{sendBlock.num != 0}$ is enabled. The transition will be taken on channel $send$, updating $\textit{sendBlock}$ values to zero. The synchronizing automaton adds the block in the $\textit{chain}$ and in $\textit{confirmedBlock}$ if it does not have that block and will process orphan blocks from $\textit{orphanBlock}$ by adding to the chain. That block will then be removed from $\textit{orphanBlock}$.

The next subsection describes the Peer automaton.

\subsection{Peer}\label{peer}
\begin{figure}[t]
  \centering
  \includegraphics[width = 10cm]{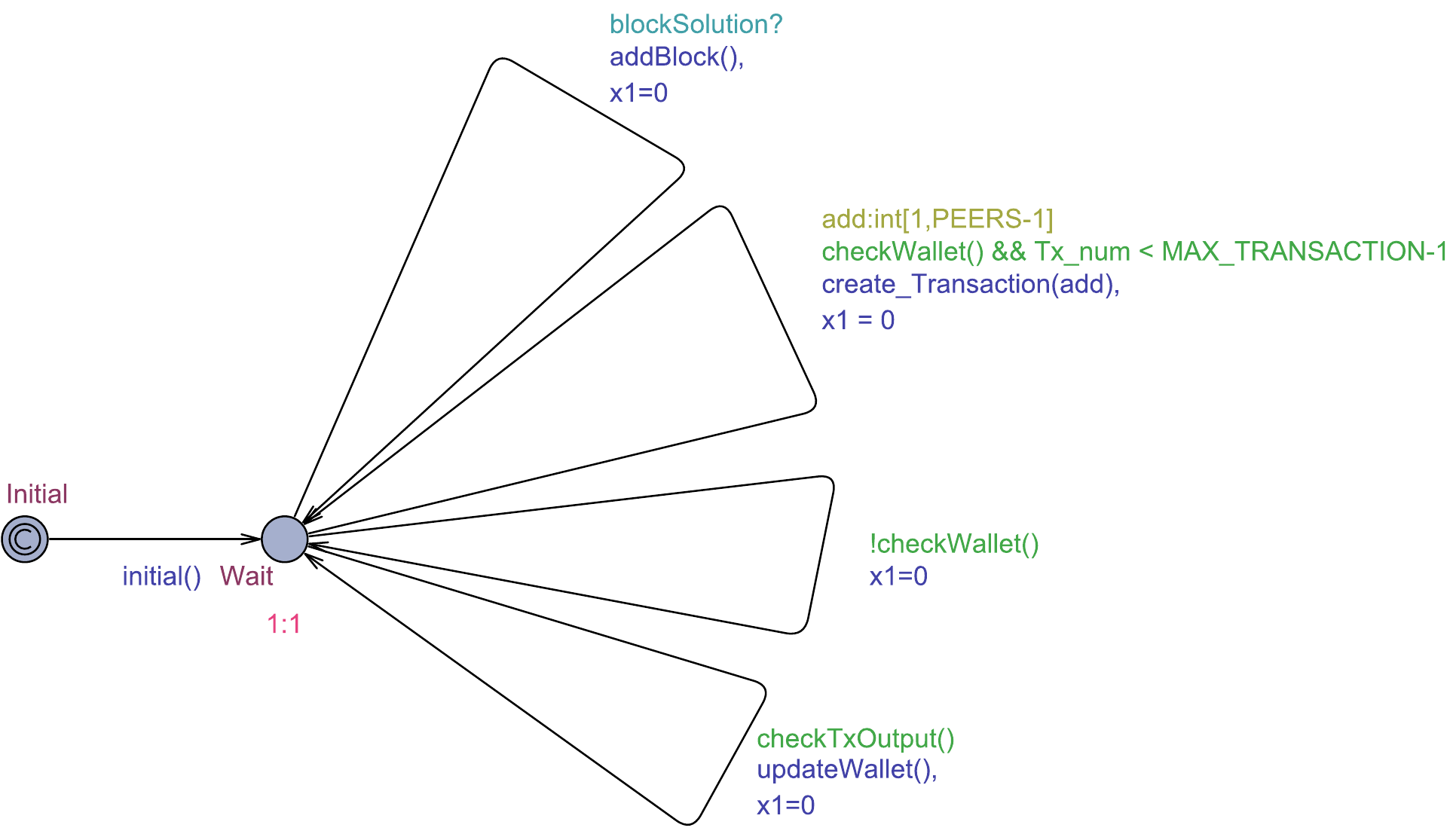}\\
  \caption{Peer automaton}\label{peertemplate}
\end{figure}

Figure~\ref{peertemplate} shows the $\textit{Peer}$ automaton. Just like pools, peers also keep track of the block-chain using variables $\textit{chain}$ and $\textit{confirmedBlock}$. The block-chain is needed for a peer to check and update the wallet if there is any transaction which has the output address as the peer's ID and it is included in the main chain. The main functionality of the peer is to create the transaction and to update the wallet. The Peer automaton has two locations: $\textit{Initial}$ and $\textit{Wait}$.

The first transition from $\textit{Initial}$ to $\textit{Wait}$ is taken to initialize the chain and the wallet. The chain is initialized as described in \ref{pool}. All peers will initially be given ten bitcoins with an initial transaction. This will be updated in the wallet. With this, peers can send bitcoins to another peer by creating transactions. The transition on creating transactions will be enabled if there are bitcoins in the wallet and if the current transaction number $\textit{Tx\_num}$ is less then the maximum transaction $\textit{MAX\_TRANSACTION}$. The select statement $\textit{add:int[0,PEERS-1]}$ allows peers to send bitcoins to any peer. The selected peer address $\textit{add}$ will be passed as parameter into the function $\textit{create\_Transaction(add)}$. This function gets the index of the wallet where a positive bitcoin value is available, searches for an empty index in $\textit{TxPool}$ and updates $\textit{TxPool}$ with the current transaction number $\textit{Tx\_num}$, the input transaction extracted from the wallet, the address of the peer (receiver), and the amount (always 1). If there is any amount left then the peer needs to send it back to itself. The function will also set the value of the transaction in the wallet to 0. This means that this peer cannot use this transaction as an input transaction in another one.

The next transition checks if there are no bitcoins in the wallet; it will be taken if there is none. To update the wallet, the condition $\textit{checkTxOutput()}$ needs to be satisfied. It checks the main chain to see if there is any transaction that has an output with this peer's ID that has not been recorded in the wallet previously. If that is the case, then this transition is taken, and the wallet is updated by function $\textit{updateWallet}$.

The last transition is on adding blocks to the block-chain using function $\textit{addBlock()}$ as described in subsection \ref{pool}.

In this model, all peers and pools are connected to each other. So there will not be any forks in the chain unless there is a malicious peer and pool. The following subsections will discuss the malicious pool and peer.

\subsection{Malicious Pool}\label{sec:mpool}

The purpose of the malicious pool in our model is to race against the network, in order to increase the length of its chain. We wanted to see whether it is possible for a malicious pool to race even when it is some blocks behind. This scenario is further illustrated in Figure~\ref{blockchainlength}. Suppose a malicious peer has spent bitcoins to some other peer and the transaction is included in block 3. This malicious peer contacts the malicious pool so that the spent bitcoins can be transferred back to this peer. The malicious pool creates a fork in the chain by adding and linking block number 7 to block number 2. Note that the malicious pool has to assign the predecessor of block 7 as block 2 since the first transaction is included in block 3. Therefore, the malicious pool has to race to make its chain longer than the chain of the honest pools. If it succeeds, then a successful double spending has occurred.

\begin{figure}[t!]
  \centering
  \includegraphics[width = 12cm]{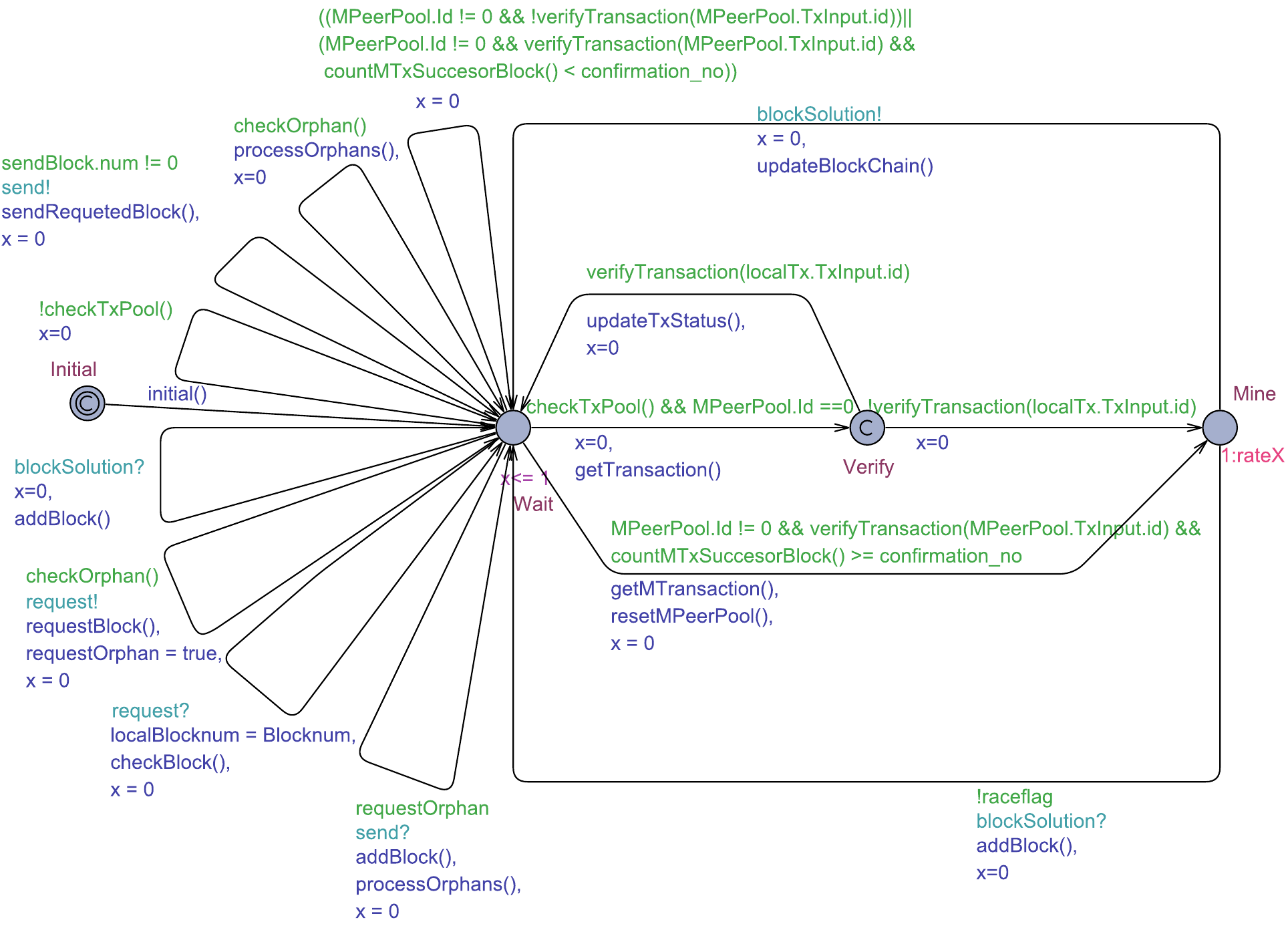}\\
  \caption{Malicious pool automaton}\label{mpool}
\end{figure}

Figure~\ref{mpool} shows the malicious pool automaton. This automaton is the same as automaton $\textit{Pool}$, except that there are two additional transitions. The first transition checks if there is any transaction from the malicious peer and whether that transaction has been included in the chain. It also checks for the number of confirmations to wait for, so that the peer to which this transaction was made, could believe that the transaction has been confirmed. The transition will be taken if the condition is true.

\begin{figure}[t!]
  \centering
  \includegraphics[width = 12cm]{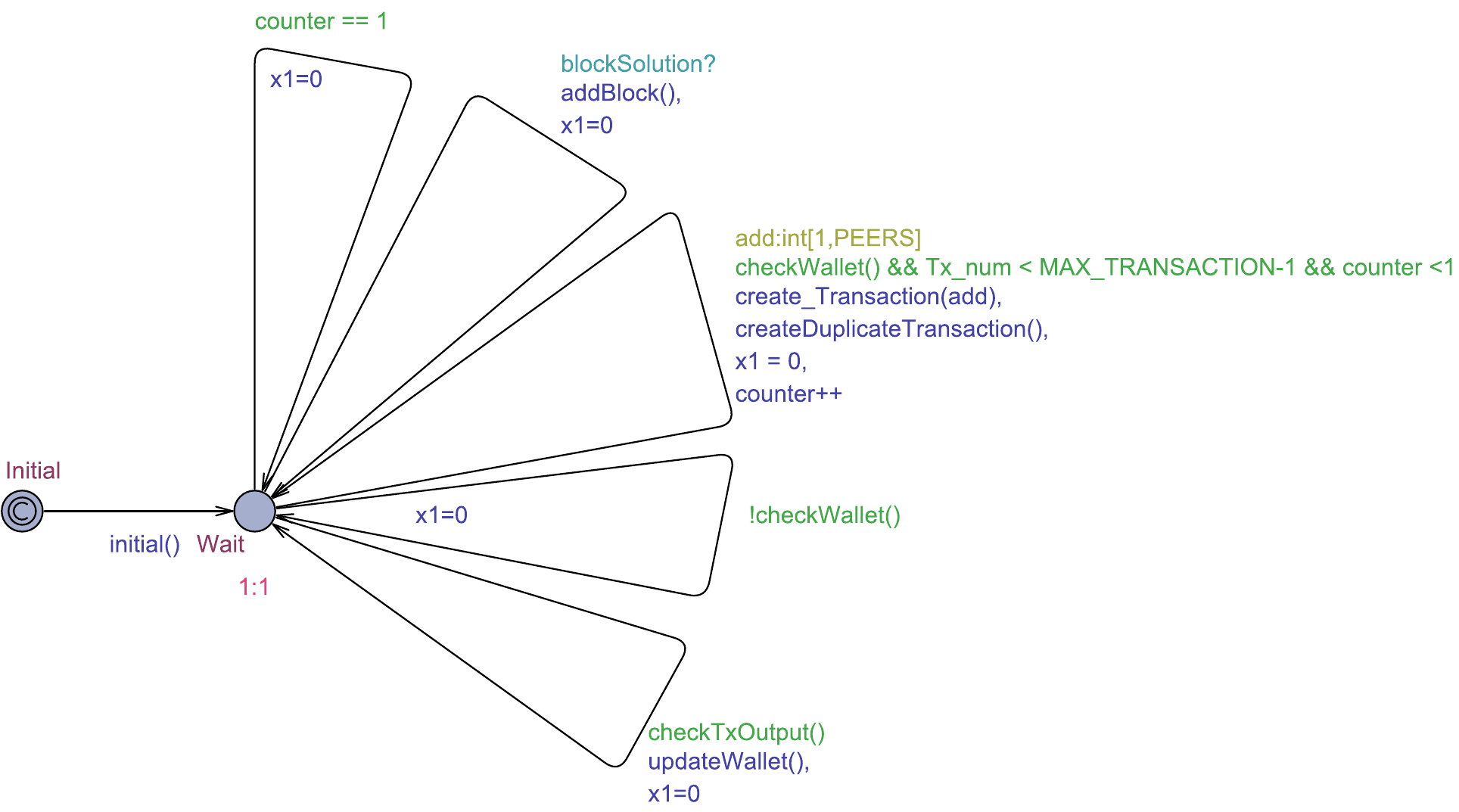}\\
  \caption{Malicious peer automaton}\label{mpeer1}
\end{figure}
The second transition is from location $\textit{Wait}$ to $\textit{Mine}$. It will be enabled if there is a transaction with the same transaction input as  the first transaction created by the malicious peer (malicious transaction) {\em and} the first transaction has been included in the chain with some blocks of confirmation. If both conditions are satisfied, then the malicious transaction will be copied into the $\textit{localTx}$ and the malicious transaction pool will be reset. At the location $\textit{Mine}$, the pool will mine a block with this transaction and will broadcast this block while transiting from location $\textit{Mine}$ to $\textit{Wait}$. The race flag $\textit{raceflag}$ will be set when there is a malicious transaction and this flag will be set to false when this malicious transaction is included in a block by this pool. This flag is used when receiving blocks at location $\textit{Mine}$. It will not allow this pool to receive any block until this transaction is included in a block. The block with the malicious transaction will be added to the chain, thus creating a fork. This fork will lead to two chains: the main chain and a side chain. Honest pools will continue building blocks on the main chain while the malicious pool will continue on the side chain. A possible result is shown in Figure~\ref{blockchainlength}. The malicious pool will also keep track of the chain built by the others in the network and will monitor when its side chain gets longer than the main chain. When that happens, the main chain for all pools will be same. Other transitions are the same as for normal pools, as described in Subsection~\ref{pool}.

\subsection{Malicious Peer}\label{mpeer}
The malicious peer will create two transactions, which will have the same transaction input. The first transaction will be placed in the transaction pool $\textit{TxPool}$, so other pools can include it in a block and proceed to the mining step. The second transaction will be given to the malicious pool.
Figure~\ref{mpeer1} shows the malicious peer automaton. One new function is introduced, $\textit{createDuplicateTransaction()}$. The purpose of this function is to create a malicious transaction and place it in a malicious transaction pool. The malicious peer automaton creates only two transactions.



\end{document}